\documentclass{aa}
\usepackage{txfonts}
\usepackage{graphicx}
\usepackage{natbib}
\usepackage{epsfig}
\newcommand{\sdss}{\emph{SDSS}}
\newcommand{\lris}{\emph{LRIS}}
\newcommand{\us}{\object{US 708}}
\newcommand{\kms}{${\rm km \, s^{-1}}$}

\newcommand{\teff}{$T_{\rm eff}$}
\newcommand{\hei}{He {\sc i}}
\newcommand{\heii}{He {\sc ii}}

\newcommand{\masy}{${\rm mas\,yr^{-1}}$}
\begin{document}

\title{\us --- An unbound hyper-velocity subluminous O star}

\author{H.~A. Hirsch\inst{1} \and U. Heber\inst{1} \and S.~J. O'Toole\inst{1}  
\and F. Bresolin\inst{2}}

\institute{Dr Remeis-Sternwarte, Astronomisches Institut der Universit\"at 
Erlangen-N\"urnberg, Sternwartstr. 7, Bamberg D-96049, Germany, \and 
Institute for Astronomy, University of Hawaii, 2680 Woodlawn Drive, 96822 
Honolulu, Hawaii USA}

\offprints{H.~A. Hirsch}

\date{Received / Accepted}

\abstract{
We report the discovery of an unbound hyper-velocity star, \us, in the 
Milky Way halo, with a heliocentric radial velocity of $+708\pm15$\,\kms. 
A quantitative NLTE model atmosphere analysis of optical spectra obtained 
with \lris\ at the Keck~I telescope
%with the Keck I telescope and the \lris\ spectrograph 
shows that \us\ is an extremely helium-rich (N$_{\rm He}$/N$_{\rm H}$=10) subluminous O type star 
with \teff$=44\,500\,K, \log g=5.23$
%Adopting the canonical mass of 0.5$M_{\sun}$  from evolution theory the 
%corresponding 
%at a distance of $19.3^{+3.1}_{-2.7}$\,kpc.
at a distance of 19\,kpc.
Its Galactic rest frame velocity is at least 751\,\kms, much higher than 
the local Galactic escape velocity indicating that the star 
is unbound to the Galaxy.
It has been suggested that such hyper-velocity stars can be formed by the 
tidal disruption of a binary through interaction with the super-massive black 
hole (SMBH) at the Galactic centre (GC).
Numerical kinematical experiments are carried out to reconstruct the path 
from the GC.
\us\ needs about 32\,Myrs to travel from the GC to its present position, 
less than its evolutionary lifetime.
Its predicted proper motion $\mu_\alpha \cos{\delta} = -2.3$\,\masy\ and 
$\mu_\delta = -2.4$\,\masy\ should be measurable by future space 
missions.
We conjecture that \us\ is formed by the merger of two helium white dwarfs 
in a close binary induced by the interaction with the SMBH in the GC and 
then escaped.

}

\titlerunning{\us\ -- an unbound hyper-velocity sdO star}
\authorrunning{H.~A. Hirsch}

\maketitle

\section{Introduction}
\label{sec:introduction}

High velocity O and B type stars at high Galactic latitudes have been known since 
decades \citep{blaa61}.
These are often called {\it runaway stars}, as they are moving away at high 
velocities from their place of birth in the Galactic plane.
None of the runaway O and B stars were known to have velocities so high as 
to exceed the Galactic escape velocity and, therefore, leave the Galaxy.
Recently, \citet{brow05} discovered a so-called hyper-velocity star (HVS), 
the faint B-type star SDSS~J090745.0+024507 (B=19.8), in the Sloan Digital 
Sky Survey (\sdss) with a heliocentric radial velocity of 853\,\kms\ unbound 
to the Galaxy.
Photometric investigations showed it to be a slowly pulsating B-type main 
sequence star \citep{fuen05}.
Soon thereafter, the $\rm 16^{th}$ magnitude star HE~0437-5439 was found to 
be a main sequence B star at a radial velocity of 723\,\kms, which exceeds 
the Galactic escape velocity (Edelmann et al., 2005).
\citet{brow05} conclude that their HVS was ejected from the Galactic centre 
(GC) because only a massive black hole could plausibly
accelerate the 3 $M_{\sun}$ main sequence B star to such an extreme velocity.
Moreover, their star's lifetime, radial velocity vector, and solar
metalicity were consistent with a GC origin.
A proper motion of $\approx 2$\,\masy\ is necessary for the star to have come 
within a few parsec of the GC \citep{gual05}, with the intrinsic proper motion
being a few tenth of a \masy\ \citep{gned05}.
HE~0437-5439, however, cannot originate from the Galactic centre unless it is a
blue straggler, because the evolutionary life time is found to be much shorter 
than the time of flight.
\citet{edel05} pointed out that HE~0437-5439 may have been ejected from the 
Large Magellanic Cloud (LMC) instead, because it is much closer to the LMC 
than to the GC and the required time of flight from the centre of the LMC
 to its present position is sufficiently short.

\citet{hill88} predicted that velocities even in excess of 1\,000\,\kms\ 
can be gained by the disruption of a binary through tidal interaction with 
the super-massive black hole (SMBH) in the GC \citep{scho03,ghez05}.
Therefore a HVS could be ejected from the GC by tidal breakup of a binary.
If HE~0437-5439 originates from the LMC centre, its existence may be evidence 
for a central massive black hole in the LMC \citep{edel05}

Here we report the discovery of a third HVS, \us, with a heliocentric radial 
velocity of $708\pm15$\,\kms.
Unlike the two known HVSs, \us\ is an evolved low mass star of spectral type 
sdO.

\section{Observations}
\label{sec:observations}

\begin{figure}
\resizebox{\hsize}{!}{\includegraphics[width=0.5\textwidth]{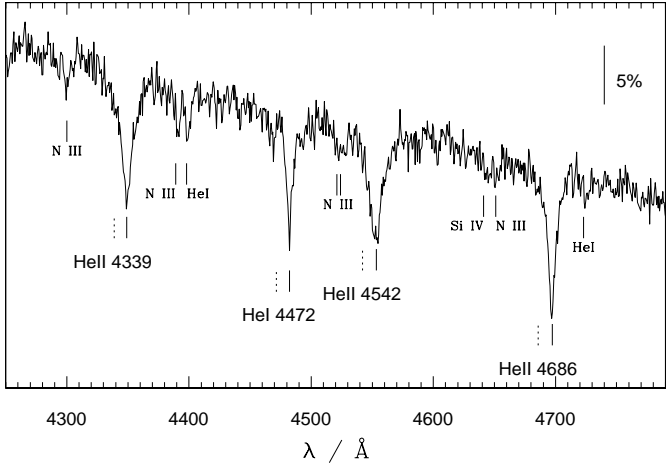}}
\label{img:us708}
\caption{Section of the spectrum of \us. Rest-wavelengths of the strongest 
lines are marked as dashed lines. Note the large redshifts.}
\end{figure}

\us\ was discovered by \citet{usher82} as a faint blue object (B=18.5) at 
high Galactic latitudes ($l=175.99^\circ, b=+47.05^\circ$), but no 
follow-up observations have been published.
The object was rediscovered by the \sdss\ as \emph{SDSS~J093320.86+441705.4} 
and $ugriz$ magnitudes of 18.35, 18.75, 19.30, 19.67, 20.05, respectively, were 
measured.

We initiated a search for subluminous O stars in the \sdss\ spectral 
database by 
selecting all objects within a colour range of 
$(u'-g') < 0.2$ and $(g'-r') < 0.1$.
%This returned a total of 11\,000 spectra.
By visual inspection we classified $\approx$100 of them as subluminous O stars 
according to their spectra showing lines of neutral as well as ionized helium.
While most stars of the sample have small radial velocities, the spectrum of 
one star, \us, is redshifted by about 10\,{\AA}.

\us\ belongs to the HesdO subclass \citep[see]{stro05} because 
no contribution of hydrogen Balmer blends to the \heii\ Pickering lines is 
visible to the eye.

%Its highly redshifted \sdss\ spectrum may indicate an unusually high 
%recession velocity.
%However, we cannot exclude that it is an artifact due to a calibration error.
However, we cannot exclude that the unusually high redshift may be an artifact.
%Therefore it deemed necessary to obtain another spectrum for verification.
For verification we used the Keck I telescope with \lris\ (Low Resolution 
Imaging Spectrometer).
The setup used was the $1\farcs5$ wide long slit with the 600/4000 grism 
for the blue arm (exposure time 900s).
% and the 900/5500 grating for the red arm.
This yields a resolution of 5\,{\AA}
%and 4.4\,{\AA} FWHM, respectively 
--- a lower resolution than \sdss, but with sufficiently high S/N to measure a 
reliable radial velocity.
%The spectrum was exposed for 900 seconds.
The spectrum was calibrated using Hg, Cd and Zn lamps for the wavelength 
range 3900\,{\AA} to 5000\,{\AA} and shows ten helium lines, out of which 
six (\hei\ 4472 \& 4922\,{\AA}; \heii\ 4339, 4542, 4686 \& 4859\,{\AA}) 
were found to be useful to measure the radial velocity.
A section of the spectrum is displayed in Fig.~1.
%\ref{img:us708}.
%The red spectrum was wavelength calibrated using the night sky lines, 
%covering 5577\,{\AA} to 6634\,{\AA}.
%There are only two lines, \hei\ 5876\,{\AA} and \heii\ 6560\,{\AA} in the red 
%channel.
%The former is blended with the interstellar \nai\ D lines due to the large 
%redshift of the stellar spectrum.
%\heii\ 6560\,{\AA} is too noisy to determine the wavelength shift reliably.
The resulting heliocentric radial velocity is $708\pm15$\,\kms, to our 
knowledge the third largest measured for any Galactic star.

\section{Atmospheric parameters and distance}
\label{sec:atmospheric}

\begin{figure}
\resizebox{\hsize}{!}{\includegraphics[height=8.0cm]{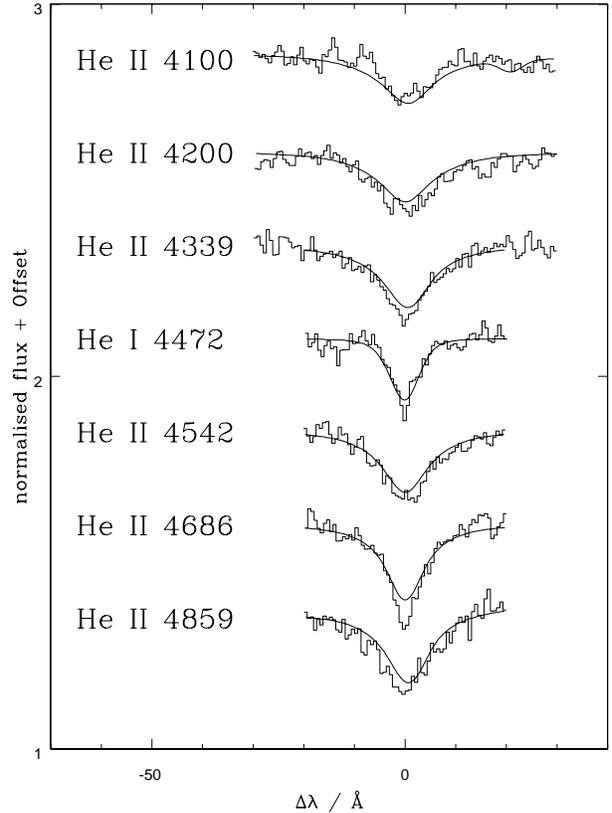}}
\label{img:us708fit}
\caption{Fit of line profiles for \us. The observed spectrum is plotted as a 
histogram, the synthetic spectrum is the solid line.}
\end{figure}

A quantitative spectral analysis was performed using an extensive grid of 
NLTE model atmospheres calculated using the latest version of the 
PRO2 code \citep{wedr99} that employs a new temperature correction 
technique \citep{drei03}.
A new detailed model atom for helium appropriate for the sdO temperature 
regime was constructed \citep{stro05}.
The model composition is hydrogen and helium only, line blanketing effects of 
H and He lines are accounted for.
To determine the stellar parameters \teff, 
$\log g$ and $\log (N_{\rm He}/N_{\rm H})$, we used the 
program \emph{FITPROF 2.2} \citep{napi99}, which uses a $\chi^2$ fit technique 
to determine all three atmospheric parameters simultaneously by matching the 
synthetic spectra to the observation.
Beforehand all spectra were normalized and the model spectra were folded with 
the instrumental profile (Gaussian with appropriate width).

The following atmospheric parameters result:
%\begin{eqnarray*}
$T_{\rm eff}$~=~45\,561$\pm$675 K, 
$\log g$~=~5.23$\pm$0.12, 
$\log (N_{\rm He}/N_{\rm H})$~=~0.99$\pm$0.18.
%\end{eqnarray*}
Note that the errors are statistical only, 
%representing the agreement between fit and observed spectrum.
The line profile fit is displayed in Fig.~2. 
Overall the fit is quite acceptable, 
%bearing in mind the S/N ratio and 
%spectral resolution.
although some line cores, \heii\ 4686\,{\AA} in particular), are not well 
reproduced (see Fig.~2).
Whether this is due shortcoming in the atmospheric models (e.g. metal line 
blanketing) or an instrumental effect remains unclear.

\begin{figure}
\resizebox{\hsize}{!}{\includegraphics[width=0.5\textwidth]{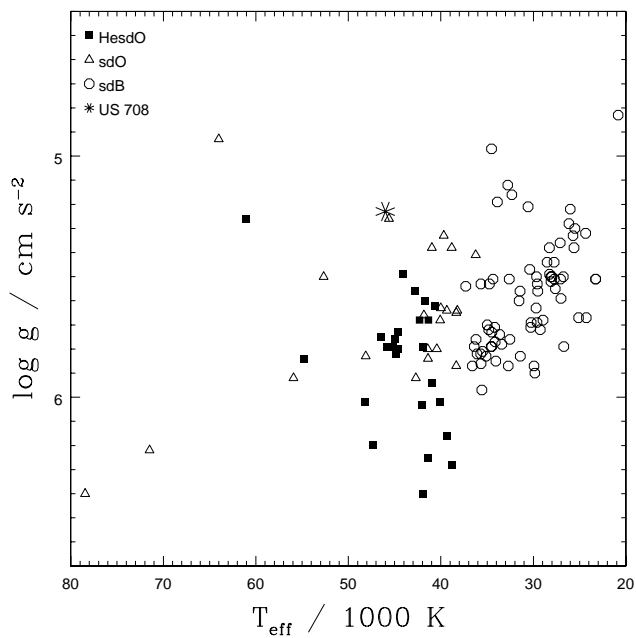}}
\label{img:tefflogg}
\caption{\teff-$\log g$-diagram of subluminous stars analyzed by
  \citet{lisk05} and \citet{stro05}. sdB stars are denoted
  by open circles, sdO stars by open triangles and HesdO by filled squares.
\us\ is marked by a star.}
\end{figure}

In Fig.~3
%\ref{img:tefflogg} 
we compare the atmospheric parameters of \us\ to
those of other HesdO, sdO and sdB stars analysed from high resolution spectra 
by \citet{lisk05} and \citet{stro05}. 
As can be seen the effective temperature of \us\
is very similar to that of most HesdO stars, whereas its gravity is slightly
lower. Nevertheless, we regard the atmospheric parameters of \us\ as typical
for sdO stars.  

In order to calculate the distance of \us\ we need to know its apparent visual 
magnitude as well as its mass.
From the star's $ugriz$ magnitudes we calculate its apparent visual 
magnitude V=$19^m.0$ using the calibration of \citet{smith02}.
For the mass we assume the canonical mass of $0.5\,M_{\odot}$ suggested by 
evolutionary models \citep{dorm93}.
Using the mass, the effective temperature, gravity and apparent magnitude, 
we derive the distance 
%as described by \citet{ram01} 
to be ${\rm 19.3^{+3.1}_{-2.7}}$\,kpc.
The star's distance from the Galactic centre is 25.8\,kpc and it is 
located 14.1\,kpc above the Galactic plane.

Correcting for the solar reflex motion and to the local standard of 
rest \citep{debi98}, the Galactic velocity components can be derived from the 
radial velocity ($U = -471$\,\kms, $V = 259$\,\kms and $W =  525$\,\kms, 
U positive towards the GC and V in the direction of Galactic rotation) 
resulting in a Galactic rest-frame velocity of 751\,\kms, indicating 
that \us\ is unbound to the 
Galaxy because the escape velocity at a galactocentric distance 
of 25\,kpc is $\approx$430\,\kms\ \citet{alsa91}.

\section{Space motion}
\label{sec:space}

Proper motion measurements are needed to reconstruct the full space 
velocity vector and trace the trajectory of \us\ back to its birth place.
The USNO-B1.0 catalog \citep{monet03} lists 
$\mu_\alpha \cos(\delta) = -6\pm1\rm\,mas\,yr^{-1}$ and $\mu_\delta = 
2\pm3\rm\,mas\,yr^{-1}$.
Given the faintness of the star, we regard the catalog errors as too 
optimistic and plausible errors to be larger than the measured components.
%However, it is possible to compute the flight time for every hypothetical origin in our Galaxy by varying the proper motion components in an iterative way.
%Since the most likely origin for a HVS is the galactic centre, we reconstructed the path of \us\ and calculated the flight time to the centre of the Milky Way.
Since the SMBH in the Galactic centre is the most likely accelerator for a 
HVS, we 
reconstructed the path of \us\ from the GC by varying the proper motion 
components.

Calculations were performed with the program ORBIT6 developed by 
\citet{odbr92}.
This numerical code calculates the orbit of a test body in the Galactic 
potential from \citet{alsa91}.
The complete set of cylindrical coordinates is integrated and positions and 
velocities are computed in equidistant time steps.
Trial values for the unknown proper motions were varied until the star 
passed through the GC with an accuracy of better than 10\,pc, 
see \citet{edel05} for details.
The resulting time of flight is 32\,Myrs with a predicted proper 
motion of $\mu_\alpha\cos(\delta) = -2.2$\,\masy\ and 
$\mu_\delta = -2.6$\,\masy. Because \us\ is the closest known HVS, its proper
motion may provide the first constraint on the shape of the Galactic potential
using HVS as proposed by \citep{gned05}.

\section{Evolution of subluminous O stars}
\label{sec:evolution}

Before discussing the origin of \us\ further, we shall discuss the 
evolutionary status of sdO stars.
They are generally believed to be closely linked to the sdB stars.
The latter have been identified as Extreme Horizontal Branch (EHB) stars 
\citep{heb86}, i.e. they are core helium burning stars with hydrogen envelopes 
too thin to sustain hydrogen burning (unlike normal EHB stars).
Therefore they evolve directly to the white dwarf cooling sequence avoiding 
the Asymptotic Giant Branch (AGB).
It may, however, be premature to assume that all the hotter sdO stars are 
descendants of the sdB stars in the process of 
evolving into white dwarfs. 
While the sdB stars spectroscopically form a homogenous class, a large variety 
of spectra is observed among sdO stars \citep{heb05}. 
Most subluminous B stars are helium deficient, whereas only a small 
fraction of sdO stars are.
Most of the latter are helium rich including a large fraction of 
helium stars (HesdO), i.e. stars for which no hydrogen Balmer line blends 
to the
\heii\ Pickering series are detectable.
\citet{heb05} summarized the results of recent spectroscopic analyses and 
provided evidence that the HesdOs are a population different from the 
sdO and sdB stars, both because of their distribution in the 
(\teff, $\log g$) diagram and their binary frequency.
While the HesdO stars cluster near \teff=45000~K, the sdO stars are widely 
spread (see Fig.3).
%~\ref{img:tefflogg}).
The fraction of sdB stars in short period binaries ($P<10d$) is high.
\citep{max01} found 2/3 of their sdB sample in be such binaries, whereas 
\citet{napi04} found a somewhat lower fraction of 40\%.
Amongst the sdO stars a similarly large fraction was found, whereas only one
HesdO star was found to be in a close binary implying a fraction of less than
5\%.
Obviously, binary evolution plays an important role for the formation of sdB 
stars and possibly also for that of the sdO stars. 
A recent population synthesis study \citep{han03} identified three channels 
to form sdB stars:
(i) one or two phases of common envelope evolution,
(ii) stable Roche lobe overflow and
(iii) the merger of two helium-core white dwarfs.
The latter could explain the population of single stars.
The simulations by \citep{han03} cover the observed parameter range of sdB 
stars but fail to reproduce their distribution in detail.
Due to the lack of binaries it may be tempting to consider 
the HesdO stars as having formed by such mergers although their distribution 
does not agree very well with the predictions from simulations \citep{stro05}. 
Neutron stars are known to
travel at extreme velocities because they are ejected by asymmetric supernova
kicks. SdO stars are remnants of low mass stars and do not suffer from supernova
explosions.

\section{The origin of \us}
\label{sec:origin}

In section \ref{sec:space} we have shown that \us\ could have originated 
from the Galactic center and reached its present position in about 30\,Myrs.
\citet{hill88} predicted that hyper-velocity stars should exist if the 
Galactic centre hosts a super-massive black hole (SMBH), because 
tidal disruption of a binary interacting with the SMBH in the centre of our 
Galaxy  can lead to ejection velocities as high as 4000\,\kms.
\citet{yutr03} confirmed this and predicted production rates of up to 
$10^{-5}$\,HVS/yr. They also explored the encounter of a single star with a 
binary SMBH and derived even higher HVS production rates.
Recently, \citet{gual05} performed three-body scattering experiments for the 
tidal disruption of a binary system by the SMBH and an encounter of a single 
star with a binary black hole and find that the the former ejects HVSs at 
higher velocities than the latter, but had a somewhat smaller ejection rate.
\citet{gual05} also explore the properties of stellar mergers in encounters 
between stellar binaries and the SMBH.
In their experiment, tailored to the properties of the HVS 
SDSS~J090745.0$+$024507, they considered equal mass 
binaries of 1.5\,$M_{\sun}$.
Merger occur only in the closest systems 
(initial separations $0.03\,AU < a < 0.05\,AU$) and only 6\% 
(in this range of separation) of the encounters result in a binary merger 
with escape of the collision product.
Hence they find this process to be very inefficient. 

The kinematical experiments presented in section 4 have shown that \us\ may 
indeed originate from the Galactic centre and travel for about 30\,Myrs to 
reach its present position in the halo.
Hence, the ejection scenarios (described above) are viable for \us.
Since the merger scenario appears attractive to explain the evolutionary 
origin of the HesdO stars, the possibility exists that the merger of a 
close binary consisting of two helium white dwarfs occurred during the 
encounter with the central SMBH.
The sdO evolutionary life time of about 100\,Myrs is consistent with its 
time-of-flight.

\section{Conclusion}
\label{sec:conclusion}

In the context of a spectroscopic study of sdO stars from the \sdss\ we 
discovered a new hyper-velocity star, \us, with a heliocentric radial 
velocity of +708$\pm$15\,\kms at a distance of about 20~kpc. 
Keck \lris\ spectra showed it to be extremely helium-rich (He/H=10 by number). 
The star is unbound to the Galaxy as its Galactic rest frame velocity 
(at least 757~\kms) exceeds, the local Galactic escape velocity 
(about 430~\kms). A kinematical experiment showed that \us\ could have been 
ejected from the Galactic center, probably by tidal interaction with the 
central super-massive black hole, if its proper motion were 
$\mu_{\alpha}\cos(\delta) = -2.2$\,\masy\ and $\mu_{\delta} = -2.6$\,\masy.
This scenario requires that the star originally was in a binary that 
was disrupted by 
tidal interaction with the SMBH and the sdO component was ejected. 
However, the binary fraction of sdO stars is very low. 
In fact, the merger of two helium-core white dwarfs is a popular scenario 
for the formation of sdO stars. If the sdO was formed by such a merger 
before interaction with the SMBH, it would require the SMBH to be a 
binary. \citet{yutr03} showed that this scenario would be quite efficient 
in producing hyper-velocity stars. However, 
the binary nature of the SMBH in the GC is purely speculative. Alternatively, 
we considered the possibility that \us, indeed may be a merger product, 
but the merger was induced by the interaction with the super-massive black 
hole in the GC and then the star escaped. 
Theoretical simulations are urgently needed to check whether tidal 
interaction between a short period binary consisting of two helium-core 
white dwarfs with the SMBH would efficiently produce mergers.

\begin{acknowledgements}
Thanks go to Norbert Przybilla for his much appreciated help with 
the Keck observations. SJOT gratefully acknowledges financial support by the 
Deutsches Zentrum f\"ur Luft- und Raumfahrt through grant 50OR0202.
\end{acknowledgements}

\end{document}